\PassOptionsToPackage{square, numbers, sort}{natbib}
\documentclass[preprint,prr]{revtex4-2}

\usepackage{amssymb}
\usepackage{graphicx}
\usepackage{amsmath}
\usepackage{color}

\begin{document}

\title{Conductive domain walls in  ferroelectrics   as tunable  coherent  THz radiation source}
\author{Ramaz Khomeriki$^{1)}$,  Kathrin D\"orr$^{2)}$,  Jamal Berakdar$^{2)}$ }

\affiliation{ $^{(1)}$Physics Department, Tbilisi State University, 3 Chavchavadze, 0128 Tbilisi, Georgia \\ 
$^{(2)}$Institut f\"ur Physik, Martin-Luther-Universit\"at, Halle-Wittenberg, D-06099 Halle/Saale, Germany}   

\begin{abstract}
THz emission  associated with currents in conductive domains in  BiFeO$_3$ following infrared  radiation is  theoretically investigated. This experimentally observed  phenomenon is explained by the  domain wall stripes  acting as metallic resonators with the oscillating charge accumulation being at the domain wall edges. The charge oscillation frequency is related to the plasma frequency inside the domain wall. The value of plasma frequency determines  both the frequency and the amplitude of the emission emanating   from the BiFeO$_3$ lattice. We show that for certain geometries of the domain wall structure and for specific polarization of the incident pulse the THz emission embodies  a non-vanishing chirality.
\end{abstract}

\maketitle
\section{Introduction}
There are a number of sources for 
THz  radiation with the basic mechanism relying on emission associated with accelerated/decaying charge current densities in various materials.\cite{Mittleman2017,Tonouchi2007,CROCKER1964,Koehler2002,PhysRevLett.53.1555,10.1063/1.100958,10.1063/1.1148382,Mittleman2017,Leitenstorfer_2023}. An example is  a relatively recent method, called spintronic THz emitter (STEs) with the emitter consisting of  ferromagnetic layer  interfaced with a  normal metal with strong spin-orbit coupling (SOC) \cite{Kampfrath2013,Seifert2016,PapaioannouBeigang}. A  
spin polarized current launched in the ferromagnet diffuses into the normal metal leading, via the inverse spin Hall effect, to   charge current burst and an associated strong broadband THz pulse.     STEs  for several materials, like CoPt alloy,  metamagnet, metal/antiferromagnetic insulator 
\cite{https://doi.org/10.1002/adma.202404174,10.1063/5.0201789,PhysRevLett.132.176703,Adam2024} as well as  engineered   STEs  
\cite{doi:10.1021/acsphotonics.1c01693,PhysRevApplied.19.L041001,doi:10.1021/acsphotonics.3c00833,PancaldiVavassoriBonetti+2024+1891+1898,ZHANG2024110418} have been demonstrated. Less studied are  analogous processes in ferroelectrics (FE) for dynamic charge current generation following IR laser irradiation. FE are integral part of data accumulation and processing devices \cite{https://doi.org/10.1002/adfm.202412332} with the advantage of being controllable via less energy-expensive probes such as electric gating and stress fields. In view of important developments in magnetic/FE compounds and phenomena \cite{RevModPhys.96.015005}, it is useful to investigate  THz emission from FE to enlarge the  material classes for THz sources. The  emitted   radiation may also give access to 
internal processes and coupling mechanisms in the sample which set the emitted radiation characteristics, as demonstrated in this work.\\
Specifically, we study theoretically the THz emission from the BiFeO$_3$ (BFO) sample depicted in Fig.\ref{fig1}. For the BFO in the stripe-domain phase, a steady-state voltage builds up due to the non-collinear  FE polarization in the neighboring domains. 
Along the few nanometer-thick domain wall (DW) the sample is conductive. The  build-in  DW voltage is induced by  internal interactions and causes band-bending at the DW area. The ferroelectric domains are much larger  ($10^3$ times) than the DW. The FE polarization in the  domains   is  homogeneous. The domains are  non-conductive but responsive, meaning they respond to electric field linearly by an amount determined by basically  the BFO dynamic susceptibility. Our  focus is not on FE switching. Rather, we are  interested in  the DW non-linear THz response at high fields, as detailed below. \\
As detailed below, for the THz emission and propagation the following  scenario is emerges.  
When irradiating the sample with an infrared  laser pulse with a central frequency below or at the (bulk)  BFO bandgap, a  charge population 
is generated   in the conduction band at the DW (the FE domains  are assumed  bulk-wise as far as electronic structure is concerned). By virtue of the build-in DW voltage,  this charge distribution is accelerated  across the DW and flows further along the domains boundaries. \\
For a sample with a metallic cap layer, such as Pt deposited on BFO, we  expect a  sizable  damping of  oscillating currents.
%
\section{Theory and modelling}
{Our aim is to develop a scheme for the emitted radiation upon launching a photo-induced charge current along the conductive DWs  while accounting for the ferroelectric dynamics in the stripe domains. Therefore, we start at first by  setting up  the relevant equation of motion of the  FE dynamics and then couple these self-consistently to the Maxwell equations. \\

We consider  a sample  that develops large area stripe domains with conductive domain walls such as BiFeO$_3/$SrRuO$_3/$DyScO$_3$ $(110)_{\rm o}$.
Since we are dealing with phenomena at large length scales (THz radiation), it is reasonable to operate within a coarse-grained approach for the FE-polarization. The stripe domains differ by the orientation of the FE polarization $\mathbf{P}_\xi$ which is aligned along either  the axis $\xi$ or $\eta$, as indicated by the arrows in Fig.(\ref{fig1})a.
For BFO films one finds net ferroelectric polarization  in the $(x,y)$ plane and out of this plane with $71^{\rm o}$ domain walls (cf.Fig.\ref{fig1}).

The  BFO  sample  of interest develops stripe domains, as indicated in   Fig.\ref{fig1}b. The spontaneous FE polarisation   is along the $-x$ axis (i.e., $[\bar 100]$ direction in Fig.\ref{fig1}).  The periodicity  allows to reduce the analysis to  a unit cell from which  the  full results is inferred.  

\section{Dielectric response} We start by examining a single FE domain with the spontaneous polarization pointing  along the $\xi$ direction. BFO is a displacive FE with  large remnant polarization. A suitable form of the  FE free energy density for  BFO in an electric field $\mathbf{E}$ reads \cite{naturecom,fpd,lde}:
\begin{equation}
{\cal F}=\alpha_1\left(P_\xi\right)^2+\alpha_{11}\left(P_\xi\right)^4+\alpha_{111}\left(P_\xi\right)^6-P_\xi E_\xi
\end{equation}
where $\alpha_1$, $\alpha_{11}$ and $\alpha_{111}$ are second, fourth and sixth order potential coefficients. By minimizing  the free-energy density functional with respect to $P_\xi$  one obtains the static (residual) polarization $P_\xi^0$.   For our purpose it is essential to capture the   (phononic) dynamical polarization  $ p_\xi(t)$  around $P_\xi^0$.  Therefore,  we write the total polarization as $P_\xi=P_\xi^0+p_\xi(t)$ and account  for linear terms in $p_\xi$  only, which results in the equation of motion  for  $ p_\xi(t)$ 
\begin{equation} \label{22}
\frac{\partial^2 p_\xi}{\partial t^2}=-\omega_0^2p_\xi+\frac{E_\xi}{\alpha_k},
\end{equation}
where $\alpha_\kappa\approx m_e a^3/e^2$ is a kinetic coefficient ($e$ and $m_e$ are the electron charge and mass and $a$ is the unit cell size), $\omega_0=\sqrt{\alpha_{FE}/\alpha_\kappa}$ and $\alpha_{FE}$ is calculated from the potential constants in harmonic  approximation, in particular, $\alpha_{FE}=2a_1+12a_{11}(P^0_\xi)^2+30a_{111}(P^0_\xi)^4$. For $\alpha_{FE}\approx 5\cdot 10^9$ Jm/C$^2$, $a\approx 0.5$nm  we obtain for the excitation frequency $\omega_0\approx 0.63$ peta Hz, which is  above the THz radiation scale of interest here. Hence, we calculate the time dependent part of polarization as a stationary solution of \eqref{22} resulting in $p_\xi=E_\xi/(\omega_0^2\alpha_k)\equiv E_\xi/\alpha_{FE}$.  The dynamic permittivity of the insulating part of the domain is  inferred  as  
\begin{equation}
\epsilon_L=\epsilon_0\left(1+\frac{1}{\alpha_{FE}\epsilon_0}\right)\approx 24 \epsilon_0   \label{epsilon}
\end{equation}
which is in line with  the reported values for the   real part of permittivity for BFO  \cite{graph}. 
\begin{figure}[hbt!]
\includegraphics[width=16cm]{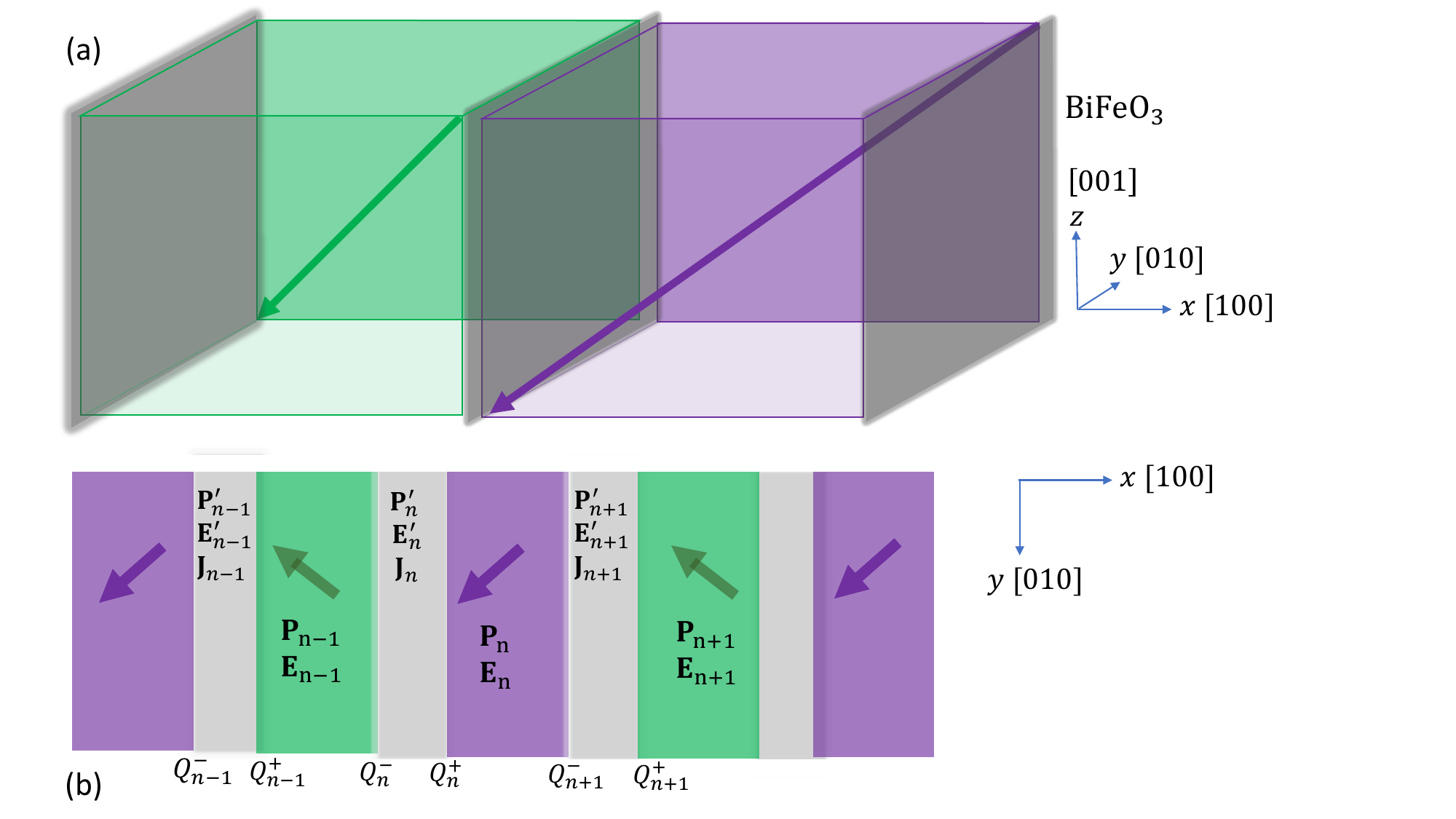}
\caption{ (a) Schematic of considered  BiFeO$_3$ (BFO)  sample in the  stripe domains (purple and green)  phase and with conductive domain walls (DW) (grey). The polarization  distribution (arrows) results in a net spontaneous polarization along the $[\bar 100]$ direction.  (b) Larger top view on the stripe domain with $x-y$ polarization components  $\mathbf{P}_n$   ($\mathbf{P}_n^\prime$) shown by arrows in the $n^{\rm th}$ stripe (DW).  Surface charges $Q^\pm_n$ and local electric fields $\mathbf{E}_n$ ($\mathbf{E}^\prime_n$) in the stripe (DW) are indicated.}
%
\label{fig1}
\end{figure}   
We note however, that  in  the case of infrared irradiation of the sample the dynamical ferroelectric response may resonate with the pulse and the values of the dynamic permittivity becomes significantly  larger than the value extracted from THz observations \eqref{epsilon}.

Having  expressed the dynamics of dielectric response by \eqref{22}, we  consider the conductive properties of ferroelectric domain structure starting from the  Drude model accounting  for the response of ferroelectric domains and domain wall
\begin{equation}
\frac{\partial {\bf J}}{\partial t}=-\frac{{\bf J}}{\tau_0}+\omega_p^2{\bf E}, \qquad \frac{\partial {\bf J}^\prime}{\partial t}=-\frac{{\bf J}^\prime}{\tau_0}+\omega_p^2{\bf E}^\prime
\label{drude0}
\end{equation} 
where ${\bf J}$, ${\bf J}^\prime$ and ${\bf E}$, ${\bf E}^\prime$ are charge current densities and electric fields inside the domains and domain walls, respectively; $\omega_p=\sqrt{e^2N_e/m_e\epsilon_0}$ is the plasma frequency and $N_e$ is the electron concentration (we assume the infrared pulse generated  $N_e$ to be the same  in the domains and the DWs), 

The scattering time  $\tau_0$ is expressible via the DC conductivity $\sigma_0$, namely  $\tau_0=m\sigma_0/e^2N_e$. Comparing  with the definition of plasma frequency $\omega_p$,  one finds  $\omega_p=\sqrt{\sigma_0/\tau_0\epsilon_0}$.  With the experimental  \cite{acs}  value $\tau_0\approx 1$ps, and  using the DC conductivity values, reported in Ref. \cite{dc} (see table 1 in Ref. \cite{dc} ), one  infers  $\sigma_0=Id/VS\approx 1.4\cdot 10^{3}\Omega^{-1}m^{-1}$, where $I$ is  the measured current, $d$ is the sample thickness, $V$ is the bias voltage and $S$ is the area associated with DW.  With the values for $\tau_0$ and $\sigma_0$  we  find for the plasma frequency  
\begin{equation}
\omega_p=\sqrt{\frac{\sigma_0}{\tau_0\epsilon_0}}\approx 4\pi\cdot 10^{12}s^{-1}.
\label{plasma22}
\end{equation}
In the subsequent explanation we  argue that this value determines the radiation frequency  of $2.1$THz that has been  detected in the experiments \cite{acs}. 

\section{ Analytic consideration}
Considering at first a BFO monodomain case, we introduce the surface charge densities ${\bf Q}^\pm={\bf Q}^\pm_0+{\bf q}^\pm(t)$ accumulated   at the opposite boundaries of the respective directions of the monodomain (${\bf q}^\pm(t)$ are their time dependent parts). Inspecting  \eqref{22} we note that the dynamics can be decomposed into two modes with different characteristics: One along $\xi$ axis (dynamical polarization $p_\xi$ along this direction is nonzero) and another mode aligned perpendicularly to $\xi$ direction (polarization $p_\perp=0$).  Assuming  the sample to be a thin film perpendicular to the $\xi$ axis (thus the depolarization factor is equal to 1 along this direction, while other factors are zero), then the electric fields components inside the domain can be written as:
\begin{equation}
E_\xi={\cal E}_\xi-(p_\xi+q_\xi)/\epsilon_0; \quad E_{\perp}={\cal E}_\perp-q_\perp/\epsilon_0. \label{xi}
\end{equation}
where ${\cal E}_\xi$ and ${\cal E}_\perp$ stand for the components of an external electric field vector, while $q_\xi$ and $q_\perp$ are dynamical surface charge densities along the boundaries of the respective directions. The electrostatic approximation is  assumed  for THz frequencies  since the size of domains are in $\mu$m range and picoseconds are sufficient for setting steady current and a quasi-static electric field distributions in the sample. Inserting \eqref{xi} into \eqref{22}, and neglecting in \eqref{22}  the time derivatives (we recall, $\omega_0$ is much larger than the characteristic THz radiation time scale) the relation follows: $E_\xi=\left(\epsilon_0{\cal E}_\xi-q_\xi\right)/\epsilon_L$,  and  from \eqref{drude0} one deduces the  equations for the current density ${\bf J}$ components by noting the relation ${\bf J}=\partial{\bf q}/\partial t$ :
\begin{eqnarray}
\frac{\partial^2 {J}_{\xi}}{\partial t^2}=-\frac{1}{\tau_0}\frac{\partial{J}_{\xi}}{\partial t}-\frac{\epsilon_0}{\epsilon_L}\omega_p^2{J}_{\xi}
+ \frac{\epsilon_0}{\epsilon_L}\omega_p^2\frac{\partial ({\epsilon_0\cal E}_{\xi})}{\partial t}, \nonumber \\
\frac{\partial^2 {J}_{\perp}}{\partial t^2}=-\frac{1}{\tau_0}\frac{\partial{J}_{\perp}}{\partial t}-\omega_p^2{J}_{\perp}
+ \omega_p^2\frac{\partial {(\epsilon_0\cal E}_{\perp})}{\partial t}.
\label{drude33}
\end{eqnarray} 
As evident,  both the frequency and the source field of the mode along the $\xi$ direction are suppressed by a large factor $\epsilon_L/\epsilon_0\gg 1$, this means that THz radiation emanating  from a mono-domain sample is polarized perpendicular to ferroelectric axis $\xi$. The radiation originating   from the mode polarized along $\xi$ is in sub-THz range and because of  the scattering time is $\tau_0\approx ps$, the mode will be damped without performing any   oscillations.  The mode with a polarization perpendicular to the ferroelectric ordering direction $\xi$ is responsible for the THz radiation with the plasma frequency $\omega_p$, as defined in \eqref{plasma22}. In reality the mono-domain sample film direction is not perpendicular to the polarization direction $\xi$, however the qualitative dynamics described by \eqref{drude33} remains correct (as confirmed by the results of numerical simulations below).

Moving on to  the THz response of FE domains with  DWs structure,  as in mono-domain case we  assume  the electric fields in the domain ${\bf E}$ and in DW ${\bf E}^\prime$ are homogeneous inside the respective parts of the structure. The matching conditions to be imposed on  the electric field induction vector ${\bf D}$ require
\begin{equation}
E^\prime_x=E_x+(p_x-p_x^\prime-q_x)/\epsilon_0; \quad E^\prime_{y,z}=E_{y,z}. \label{charge}
\end{equation}
 ${\bf p}^\prime$ stands for the ferroelectric polarization inside the domain wall. As shown below, one finds  $|{\bf p}^\prime|\gg |{\bf p}|\gg\epsilon_0|{\bf E}|$, and therefore the emission is dominated by the current density  dynamics inside  the domain walls along the $x$ direction,  meaning, the current inside the DWs along the  $x$ direction is much larger than other currents (including those along  $y$ and $z$ directions inside DWs as well as  all directions inside FE domains). This direction of the dominant  current density  determines the THz radiation, while the radiation coming from other parts of the structure is marginal. 

\section{ Numerical simulations} Let us start again with the mono-domain, thin film consideration. Let the  film be  in the $xy$ plane and the polarization is along the  $xyz$ pseudo-cube diagonal direction (see Fig. 1).  The dynamical equations for the time dependent part of polarization vector and the current density components are  
\begin{eqnarray}
&&\frac{\partial^2 p_\xi}{\partial t^2}=\frac{E_x+E_y+E_z}{\sqrt{3}\alpha_k}-\omega_0^2p_\xi, ~~ \frac{\partial^2 {\bf q}}{\partial t^2}=\omega_p^2{\bf E}-\frac{1}{\tau_0}\frac{\partial {\bf q}}{\partial t} \nonumber \\
&&E_{x,y}={\cal E}_{x,y}-q_{x,y}/\epsilon_0, \quad E_z={\cal E}_z-q_z/\epsilon_0-p_\xi/\sqrt{3}\epsilon_0 ~~~~
\label{monodomain}
\end{eqnarray}
which  assumes that the  depolarizing factor is nonzero only along the  $z$ axis. Expressions \eqref{monodomain} form a close set of equations for simulating the current density  and the polarization dynamics in BFO mono-domain thin film. First, we use the  infrared pulse external field ${\boldsymbol{\cal E}}$ with $s$ polarization (along $y$  direction). The results are displayed in Fig. \ref{Rmono}(a). As expected from the structure of Eqs.(\eqref{monodomain}), in addition to  the  $y$ component of the induced current density,  a $z$ component is also excited, while the $x$ component remains zero. On the other hand, irradiating the sample with $p$ polarized light ($xz$ polarization) with the same intensity yields the  current density distribution displayed in Fig. \ref{Rmono}(b). All the frequencies are close to 2.1 THz. The asymmetry character of $z$ component current density (cf. Fig. \ref{Rmono}(b)) is caused by a damped lower frequency mode predicted by the analytic considerations of the  previous section. 
\begin{figure}[hbt!]
\includegraphics[width=16cm]{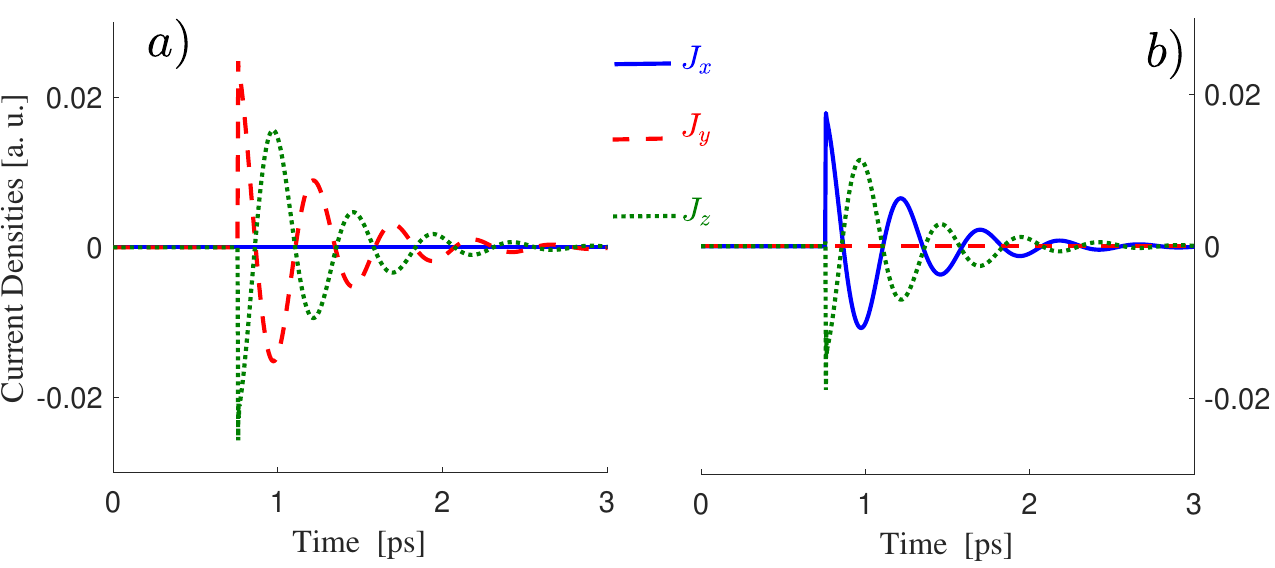}
\caption{Results of numerical simulations on the set of equations \eqref{monodomain}. In (a) the currents in the monodomain film lying in $xy$ plane in case of illumination of $s$ polarized pulse (with electric field along $y$ axis) are shown. In graph (b) the pulse is polarized in incidence  ($xz$) plane (i.e., $p$ polarization). }
\label{Rmono}
\end{figure}

In case of multiple domains structure, the equations of motion read
\begin{eqnarray}
&&\frac{\partial^2 p_\xi}{\partial t^2}=\frac{E_x\pm E_y+E_z}{\sqrt{3}\alpha_k}-\omega_0^2p_\xi, ~~ \frac{\partial^2 {\bf q}}{\partial t^2}=\omega_p^2{\bf E}-\frac{1}{\tau_0}\frac{\partial {\bf q}}{\partial t} \nonumber \\
&&E_{x,y}={\cal E}_{x,y}, \quad E_z={\cal E}_z-q_z/\epsilon_0-p_\xi/\sqrt{3}\epsilon_0. 
\label{domain}
\end{eqnarray}
In the first equation  the sign "$\pm$" accounts for the different domains. Note,    charge with different signs accumulated at the domain wall boundaries  has no effect on the domain. Along  the $y$ direction the sample is large in size such that in THz scale, there will be no quasi-static distribution of charges at the boundaries. 

 Inside the domain wall  interface effects are crucial and are determined by an interplay between the
 inhomogeneous distribution of dipoles and the  free charges.  The local field acting on the domain wall dipoles amounts  to the external electric field ${\boldsymbol{\cal E}}$, while the free charges inside the domain wall feel a "macroscopic" electric field created by domain wall dipoles, free charges at the domain wall boundaries, and  the electric field due to the polarized domains. Moreover,    the domain wall interface dipoles have additional degree of freedom, namely they can rotate around $[10\,-\!$1] axis,  as illustrated  in Fig. \ref{RDW} (a) and have smaller oscillation frequency $\omega_0^\prime$  than the one for dipoles inside the domain  $\omega_0$ (for dipole rotatory motion equations please see \cite{PhysRevB.109.045428} ). The kinetic parameter  $\alpha_k$ is  the same. Thus, we can write the equation of motion  for the dipoles and free charges inside the domain wall as follows:
\begin{eqnarray}
&&\frac{\partial^2 {\bf p}^\prime}{\partial t^2}=\frac{{\bf e}_{P_0}\times{\bf e}_{P_0}\times{\boldsymbol{\cal E}}}{\alpha_k}-\left(\omega_0^\prime\right)^2{\bf p}^\prime, \nonumber \\ 
&&\frac{\partial^2 {\bf q^\prime}}{\partial t^2}=\omega_p^2{\bf E}^\prime-\frac{1}{\tau_0}\frac{\partial {\bf q}^\prime}{\partial t},\label{eqdw}  \\
&&E_{x}^\prime={\cal E}_{x}-q_x^\prime/\epsilon_0-\left(p_x^\prime-p_\xi/\sqrt{3}\right)/\epsilon_0, \quad E_{y}^\prime={\cal E}_{y},
\nonumber \\
&&E_z^\prime={\cal E}_z-\left(q_z^\prime+q_z\right)/\epsilon_0-p_\xi/\sqrt{3}\epsilon_0. \nonumber
\end{eqnarray}
The first equation   accounts for the rotatory motion of the interface dipoles, where ${\bf e}_{P_0}$ stands for the unit vector along $[10-\!1]$ axis. 

\begin{figure}[hbt!]
\includegraphics[width=8cm]{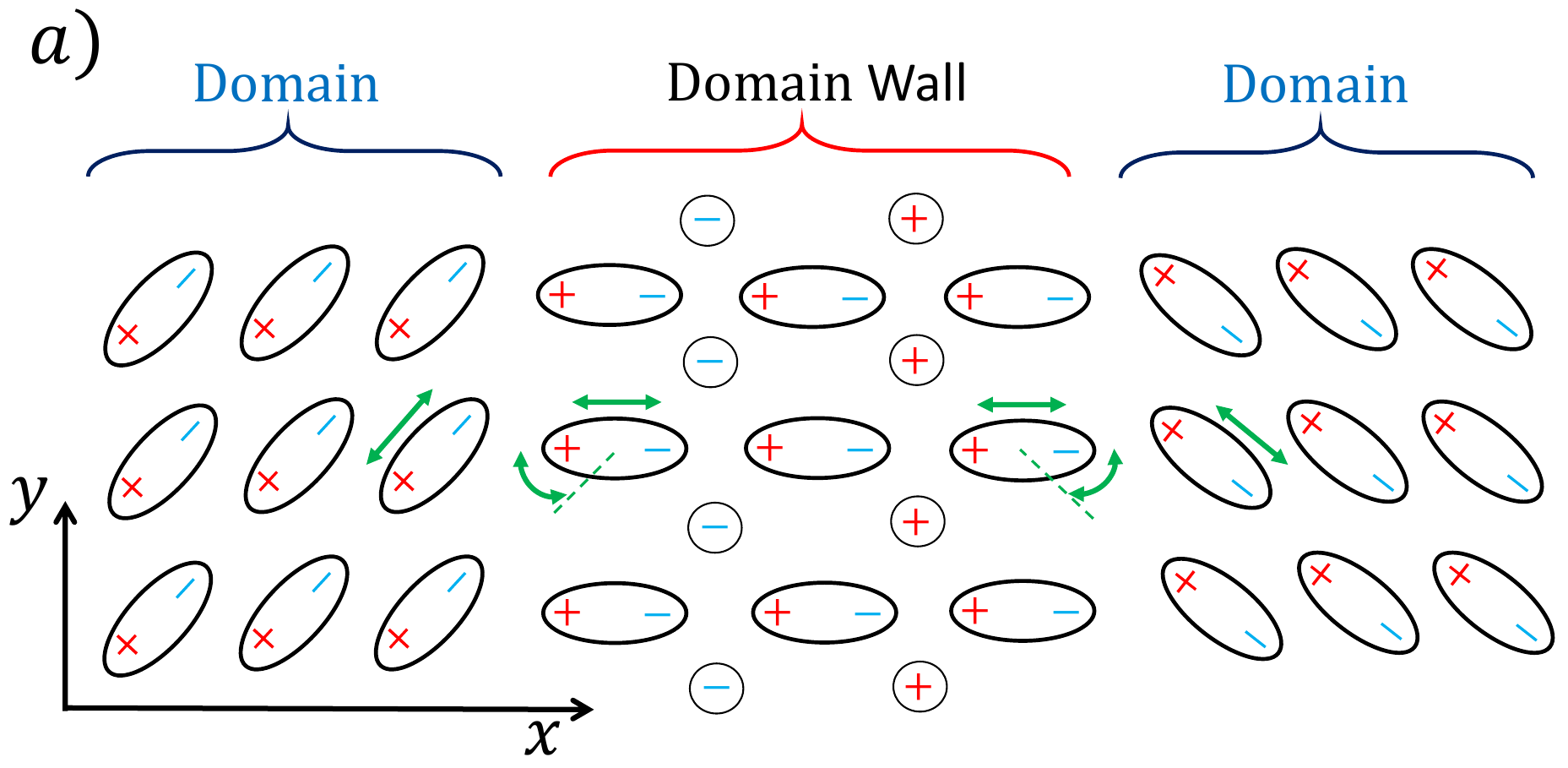}
\centering \includegraphics[width=8cm]{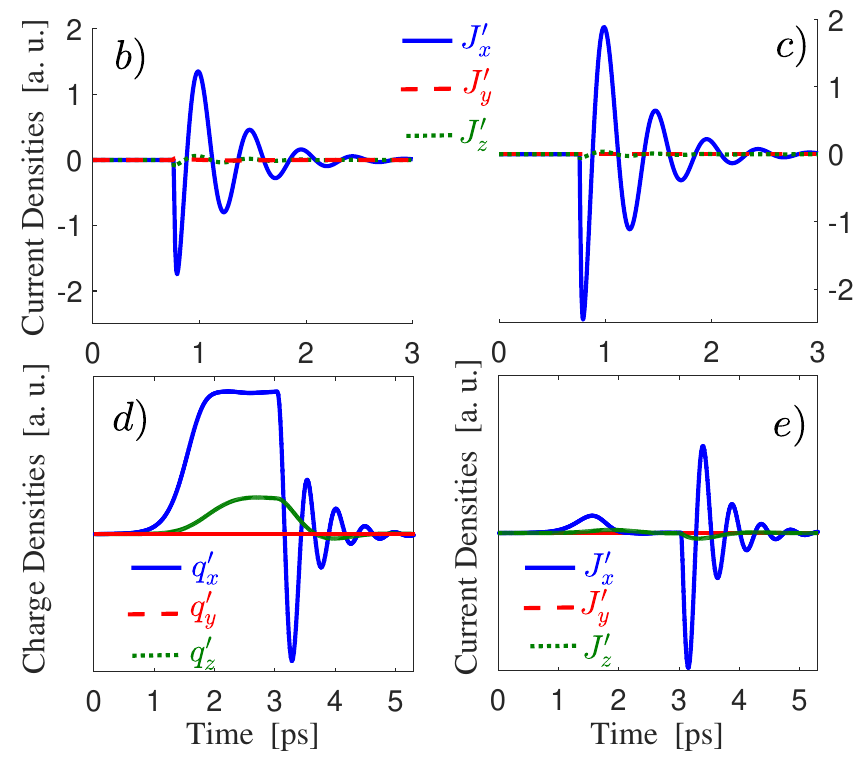}
\caption{(a)  Schematics of the distribution of dipoles in domains, domain walls, and their interface, shown as 
a projection on the $xy$ plane while the  dipoles extend along $z$ direction  (in accordance with  Fig. 1(a)). Green arrows indicate oscillation and rotational directions of interface dipoles. The disposition of free charges are indicated by circles. (b) and (c) shows the current density oscillations inside domain walls according to numerical simulations on the set of equations \eqref{domain} and \eqref{eqdw} irradiating with $s$ and $p$ polarized infrared light pulses, respectively.  (d) and (e)  show the evolution of excess charge and current densities inside domain walls during  gradual charging and sudden discharge of the domains structure.}
\label{RDW}
\end{figure}

 Figs. \ref{RDW}(b) and (c)  show simulation results for current density oscillations inside the domain walls for $\omega_0/\omega_0^\prime=3$.
 We observe  dominating current oscillations along $x$ inside the domain wall while currents in domain wall along $z$  direction are much smaller. Beside that, the current oscillations in the domains are several orders of magnitude smaller, in line with
 the analytical estimates of  the previous section. 
 Comparing the current densities in the domain wall with the currents inside monodomain (see Figs. \ref{Rmono} and \ref{RDW}) while 
 considering  the different areas of domain wall and monodomain, we conclude  that the radiation emanating  from the single domain wall and whole mono domain are comparable, in full accordance with experimental observations of Ref.\cite{acs}.

As basically  only currents along the $x$ direction are excited, the THz radiation stemming  from domain wall structure is linearly polarized irrespective of incident pulse  polarization. The radiation  characteristics becomes more involved 
for  noncollinear domain wall which can be in practice be exploited to  infer structural information on the domain walls by polarization analysis. For  example,   one may observe chiral  emission: The electric ${\bf E}$ and magnetic ${\bf H}$ fields as well as  the  chirality $\chi$ are evaluated  in far-field regions   as
\begin{eqnarray}
{\bf E}\sim \sum\limits_n\left[\left[{\bf J}_n\times{\bf e}_r\right]\times{\bf e}_r\right], \quad
{\bf H}\sim \sum\limits_n\left[{\bf J}_n\times{\bf e}_r\right],
\nonumber \\ 
\chi=\frac{1}{c^2}\left({\bf H}\cdot\frac{\partial \bf{E}}{\partial t}-{\bf E}\cdot\frac{\partial \bf{H}}{\partial t}\right). \label{chirality}
\end{eqnarray}
 $n$ numbers the domain walls and ${\bf J}_n$ is the associated current. One can manipulate the emitted THz chirality by adjusting the polarization of femtosecond laser pulses on the domain wall structure.
}
\begin{figure}[hbt!]
\includegraphics[width=14cm]{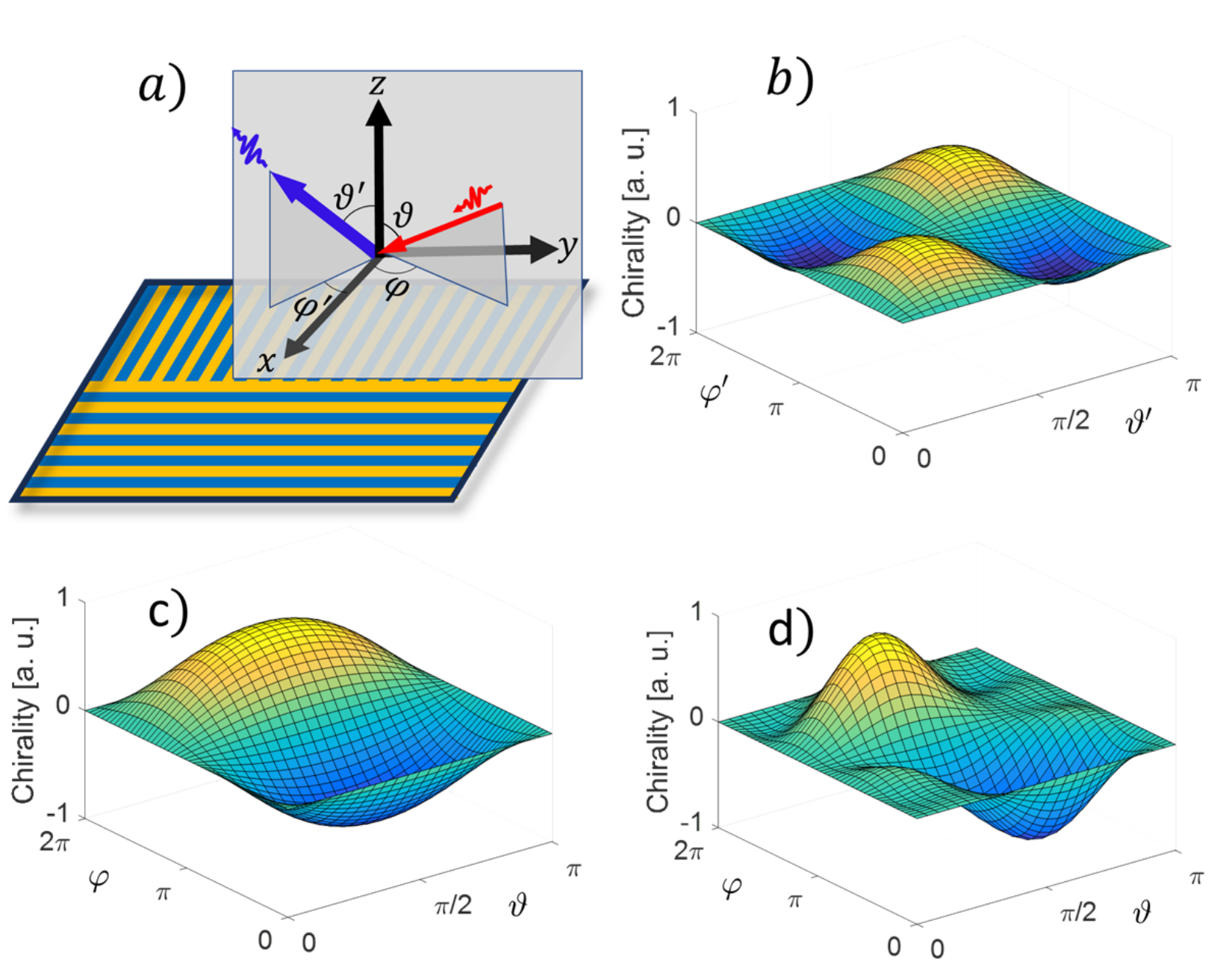}
\caption{(a)  Schematics for the THz emission from the noncollinear FE domains with conductive DWs, inset shows the scattering and reflected angles notations. Red (blue) arrow indicates the IR (THz) electric field propagation direction. In b) a circularly polarized infrared pulse is incident  from the top, along the $z$ axis and the graph shows the chirality density distribution of the radiated THz signal. In c) the infrared  pulse is $s$ polarized  and propagates along the angles $\vartheta$ and $\varphi$ while the   chirality density of the emitted THz field is monitored along $z$ axis. d) is the same as graph c), but  the incoming IR pulse is $p$ polarized. }
\label{pic}
\end{figure}

\section{Chiral emission from conductive DWs}
An advantageous feature of ferroelectric DWs is that they can be engineered, e.g. as to have anisotropic optical properties, or can be   manipulated externally , e.g. locally  by a scanning tip, gating or  strain \cite{LIU2022100943,doi:10.1021/acs.nanolett.6b04512,doi:10.1021/acsami.1c22248,doi:10.1021/acs.nanolett.3c01638,https://doi.org/10.1002/aelm.202300640}. This renders possible   steering of  emission properties 
by   applied external probes and/or using the respective samples. As far as THz emission is concerned, conductive FE DWs can be  utilized  as  phase-change optical materials  or as candidates for spatiotemporal dielectric response.  As an example, we consider the 
the fabricated BFO sample with non-collinear stripe domains structure which has been reported in \cite{ZHENG2022274}.  The simulations are done based on the equations presented in the previous sections.  We  perform the calculations for each stripe orientation separated and sum up coherently the resulting fields.\\
 Fig. \ref{pic}(a-d) demonstrate  one of  emergent new features by considering the chirality density, here  it is related to the local polarization states of the fields. 
 For a sample size of $30\times30\mu$m and  various scenarios for the 
 incident pulse direction, as  set  by the 
 scattering polar and azimuthal angles $\vartheta$ and $\varphi$, respectively (Fig.(\ref{pic}a).  
 The angles of 
 reflected THz radiation  are denoted by $\vartheta^\prime$ and $\varphi^\prime$.  
 While for the collinear case we find mainly linearly polarized emission, here the chirality density pattern in Fig. \ref{pic}(b-d) proves that we can  control the chirality density by the polarization state of the IR pulse and/or its incident direction and  produces so at certain angles of emission fully polarized THz emission. 

\section{Conclusions}
In summary, our theoretical considerations and simulations  indicate  the  significant potential of conductive DWs in ferroelectric BFO  as a source of 
THz emission.  Depending on the 
preparation conditions and/or external probes, the emission characteristics can be controlled. As an example,  samples with noncollinear domains  under  infrared irradiation  emit THz fields with a finite chirality density distribution that can be  manipulated by changing the polarization and angle of scattered infrared pulses. 

\acknowledgments This research has been funded by the DFG through B06 and A05 within the SFB-TRR227.
We thanks Tom Seifert and Tobias Kampfrath for discussions and suggestions. 
\bibliography{APL,STE}

\end{document}